\documentclass[aps,preprint,showpacs,groupedaddress,floatfix]{revtex4-1}
\usepackage{graphicx}
\usepackage[dvips,usenames]{color}
\usepackage{txfonts}
\usepackage{units}
\usepackage{bm}

\newcommand{\uK}{\ensuremath{\mu{\mbox{K}}}}

\newcommand{\micron}{\ensuremath{\mu{\mbox{m}}}}

\begin{document}

\title{Finite range corrections near a Feshbach resonance and their role in the Efimov effect}

\author{P.~Dyke, S.~E.~Pollack, and R.~G.~Hulet, }
\affiliation{Department of Physics and Astronomy and Rice Quantum
Institute, Rice University, Houston, Texas 77005, USA}

\date{\today}

\begin{abstract}
 We have measured the binding energy of $^7$Li Feshbach molecules deep into the non-universal regime
 by associating atoms in a Bose-Einstein condensate with a modulated magnetic field.  We extract
 the scattering length from these measurements, correcting for non-universal short-range effects using
 the field-dependent effective range. With this more precise determination of the Feshbach resonance
 parameters we reanalyze our previous data on the location of atom loss features produced by the Efimov effect \cite{PollackSci09}.  We find the  measured locations of the three- and four-body Efimov features to be consistent with universal theory at the 20-30\% level.

\end{abstract}

\pacs{03.75.Kk,03.75.Lm,47.37.+q,71.23.-k}

\maketitle



Efimov showed more than 40 years ago that three particles interacting via resonant two-body
interactions could form an infinite series of three-body bound states as the two-body $s$-wave
scattering length $a$ was varied \cite{Efimov70}. In the limit of zero-range interactions, the
ratios of scattering lengths corresponding to the appearance of each bound state were predicted to
be a universal constant, equal to approximately 22.7.  The only definitive observations of the
Efimov effect have been in ultracold atoms, where the ability to tune $a$ via a Feshbach resonance
\cite{Duine04, ChinRMP10} has proven to be essential.  Since the first evidence for Efimov trimers
was obtained in ultracold Cs \cite{Kraemer06}, experiments have revealed both three- and four-body
Efimov states in several atomic species.  Although the Efimov effect has now been confirmed,
several open questions remain, including a full understanding of the role of non-universal finite
range effects.  Accurate comparisons with theory require that these non-universal contributions be
quantitatively determined and incorporated.

We previously characterized the $F=1, m_F=1$ Feshbach resonance in $^7$Li, which is located at
approximately 738 G, by extracting $a$ from the measured size of trapped Bose-Einstein condensates
(BEC) assuming a mean-field Thomas-Fermi density distribution \cite{PollackPRL09, PollackSci09}.
These data were fit to obtain $a(B)$, the function giving $a$ vs.~magnetic field, which was used to
assign values of $a$ to Efimov features observed in the rate of inelastic three- and four-body loss
of trapped atoms \cite{PollackSci09}.  More recently, two groups have characterized the same
Feshbach resonance by directly measuring the binding energy, $E_\mathrm{b}$, of the weakly-bound
dimers on the $a > 0$ side of the Feshbach resonance \cite{Gross10,Gross11,Navon11}.  These
measurements disagree with our previous measurements based on BEC size.  The disagreement in the
parameters characterizing the Feshbach parameters is sufficiently large to affect the comparison of
the measured Efimov features with universal theory.

In this paper, we report new measurements of $E_\mathrm{b}$, which we fit to obtain $a(B)$.  The
measurement of $E_\mathrm{b}$ has fewer systematic uncertainties than the BEC size measurement,
which is affected at large scattering length by beyond mean-field effects and by anharmonic
contributions to the trapping potential.  The extraction of $a$ from $E_\mathrm{b}$ can therefore
be more accurate, and unlike the condensate size measurement, $E_\mathrm{b}$ is related to $a$ for
both thermal gases and condensates.  We have measured $E_\mathrm{b}$ far enough from the Feshbach
resonance that $E_\mathrm{b}$ no longer depends quadratically on the detuning of $B$ from
resonance, as it would in the universal regime \cite{Duine04,ChinRMP10}.  We show that a commonly
adopted correction for non-universal finite range effects, which depends on a single value for the
effective range, does not fit the data as satisfactorily as more complex two-channel models
\cite{Lange09, Prico11} that accommodate a field-dependent effective range, or a model that
incorporates an explicit calculation of the effective range.  We employ the latter strategy to
produce an improved $a(B)$ function to reanalyze our three- and four-body loss data to obtain more
accurate locations of the Efimov features.

Our experimental methods for producing BECs and ultracold gases of $^7$Li have been described in
detail previously~\cite{PollackPRL09}. Atoms in the $|F=1, m_F=1\rangle$ state are confined in an
optical trap formed from a single focused laser beam with wavelength of $1.06\,\micron$.  A bias
magnetic field, directed along the trap axis, is used to tune $a$ via the Feshbach resonance. For
the new data presented here the axial and radial trapping frequencies are 4.7\,Hz and 255\,Hz,
respectively. We adjust the magnetic field to give $a \sim 200\,a_0$, where $a_0$ is the Bohr
radius, and use forced evaporation to produce either ultracold thermal clouds with temperatures of
$\sim$$3\,\uK$, or condensates with a condensate fraction that we estimate is greater than $85\%$.
We then adiabatically ramp the field to the desired value and employ \emph{in situ} imaging, either
polarization phase-contrast \cite{Bradley97} when the density is high, or absorption imaging in
less dense clouds.

Atoms are associated into Feshbach molecules by resonantly oscillating the magnetic field at a
frequency $h\nu_\mathrm{mod} = E_\mathrm{b} + E_\mathrm{kin}$, where $E_\mathrm{b}$ is taken to be
positive for a bound state, and $E_\mathrm{kin}$ is the relative kinetic energy of the atom pair
\cite{Thompson05, Hanna07}.  The weakly-bound dimers formed in this way are lost from the trap
through collisional relaxation, presumably into deeply-bound vibrational levels \cite{Thompson05}.
This technique has been used in studies of both homonuclear \cite{Thompson05, Gaebler07, Lange09}
and heteronuclear \cite{Papp06, Zirbel08, Weber08} Feshbach resonances, in addition to the specific
hyperfine state of $^7$Li studied here \cite{Gross10, Gross11, Navon11}.

The oscillating field is produced by a set of auxiliary coils that are coaxial with the bias coils
producing the Feshbach field. The amplitude of this field ranges from 0.1\,G to 0.6\,G and the
duration of modulation ranges from 25\,ms to 500\,ms depending on magnetic field. The number of
remaining atoms are measured as a function of the frequency of the oscillating field.  We find that
in the case of a BEC the loss spectra are fit well by a Lorentzian lineshape, while for a
thermal gas, we fit the loss spectra to a Lorentzian convolved with a thermal Boltzmann distribution
\cite{Hanna07, Weber08, Gross10}.

\begin{figure}
\includegraphics[angle=-90,width=1.0\columnwidth]{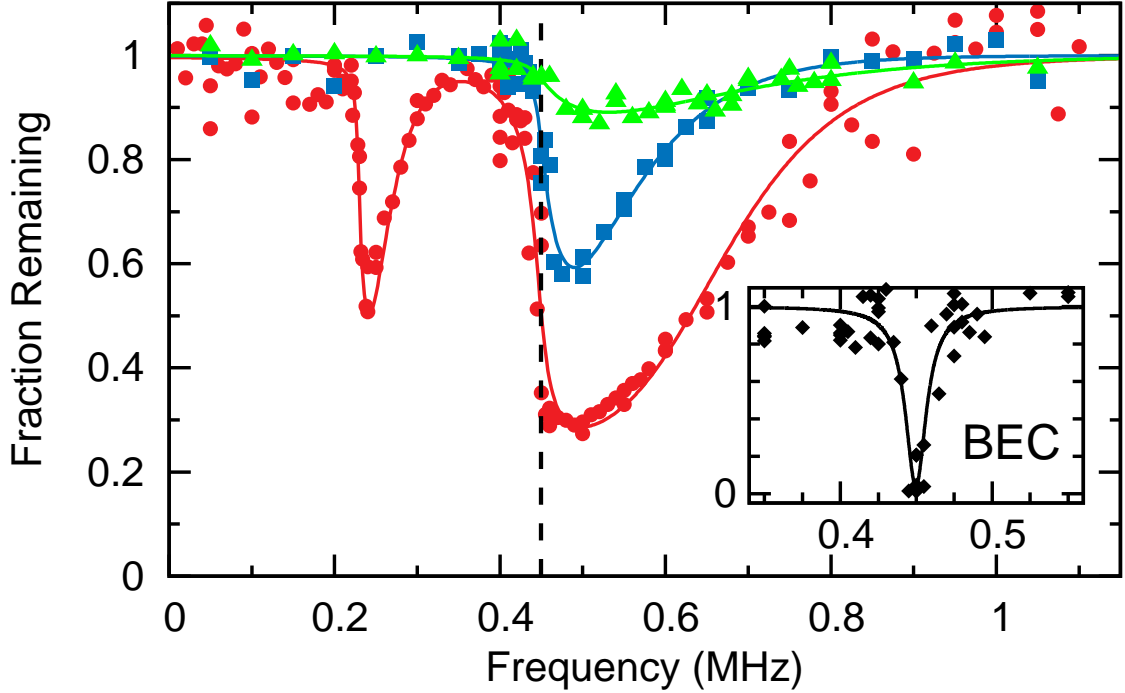}
\caption{(Color online)  Magneto-association induced loss at $B = 734.5\,$G, where $a \simeq
1100\,a_0$. The main plot shows loss spectra for thermal gases with the following modulation amplitudes and approximate temperatures: ({\color{OliveGreen} $\blacktriangle$})~0.57\,G\,/\,$10\,\uK$,
({\color{RoyalBlue} $\blacksquare$})~0.14\,G\,/\,$3\,\uK$, and ({\color{Red}
\large$\bullet$})~0.57\,G\,/\,$3\,\uK$. The solid curves are fits to Lorentzians convolved with
thermal Boltzmann distributions. For the 0.57\,G\,/\,$3\,\uK$ data the lower frequency resonance is a subharmonic response, while the primary resonance is thermally broadened by strong modulation.
The inset ({\color{Black} $\blacklozenge$}) corresponds to a BEC
with a modulation amplitude of 0.14\,G.  The solid black line is a Lorentzian fit to the condensate
resonance and the vertical dashed line in the main figure is the resonance location $E_\mathrm{b} /
h = 450\,$kHz found from this fit.} \label{fig:bindingloss}
\end{figure}

Figure \ref{fig:bindingloss} shows characteristic loss spectra at 734.5\,G (where $a \simeq
1100\,a_0$) for several different temperatures and modulation amplitudes. The Lorentzian component
fits to a linewidth of 8\,kHz, which provides a lower bound on the lifetime of the molecular state
of $20\,\mu$s.
There is no
systematic shift in the resonance location with temperature or modulation amplitude, but for large
amplitude modulations and sufficiently low temperature we observe a nonlinear resonance at
$\frac{1}{2} E_\mathrm{b}/h$.  No other subharmonics are seen.  A similar nonlinear response was
reported previously \cite{Weber08}.

\begin{figure}
\includegraphics[width=1\linewidth,bb=9 168 517 585]{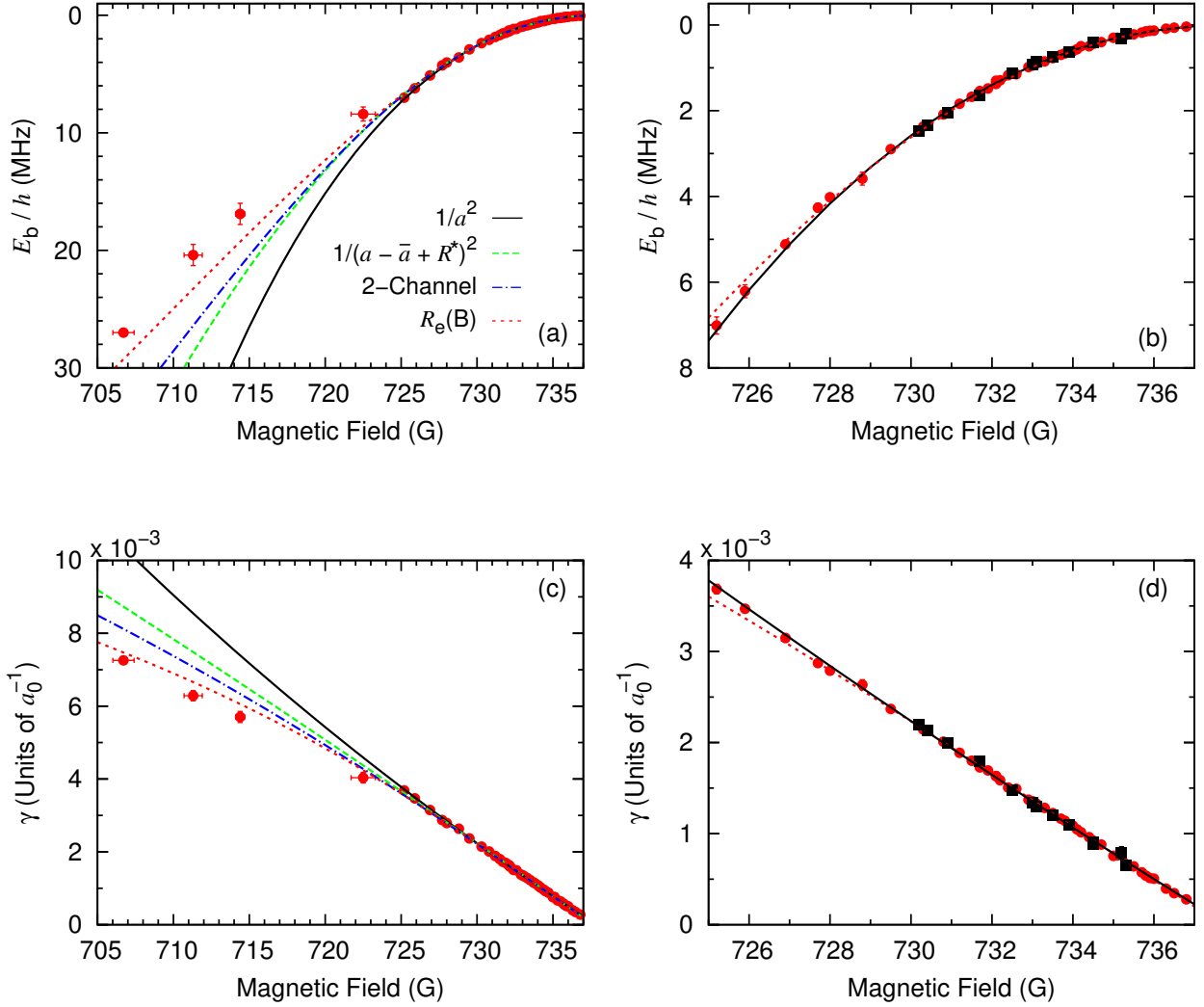}
\caption{(Color online) Results of modulation spectroscopy using condensates ({\color{Red} $\bullet$}) and $\sim$$3\,\uK$ thermal clouds ({\color{Black} $\blacksquare$}).
(a) and (b): $E_\mathrm{b}$ vs.~$B$,
and (c) and (d): the same data plotted as $\gamma \equiv (mE_\mathrm{b}/\hbar^2)^{1/2}$ vs.~$B$. The vertical error bars correspond to uncertainty in fitting to the binding energy resonances, while the horizontal error bars are the statistical uncertainties due to shot-to-shot variations of the magnetic field.  The relatively large error bars below 725\,G arise from the broadening of the resonance from the strong modulation required to produce a detectable signal.
The lines are fits of the measurements to Eq.~(\ref{eqn:fr}) using the various models.
Solid (black): universal model, $E_\mathrm{b} = \hbar^2/ma^2$;
dashed (green): simple two-channel model, Eq.~(\ref{eqn:binding}), with the parameters given in the text; dot-dashed (blue): complex two-channel model given in Ref.~\cite{Prico11} (Eq.~26), again with
parameters as given in the text (the two-channel model of Ref.~\cite{Lange09} gives similar
results); and dotted (red): Eq.~(\ref{eqn:gamma}) using $R_\mathrm{e}$($B$) from
Fig.~\ref{fig:FB_RE}.  The fits exclude data below 725\,G, where $a
\simeq 250~a_0$ is no longer much greater than $R^*$, and the validity of the non-universal
corrections becomes questionable.  While the nominal fitting parameters are $a_\mathrm{bg}$, $\Delta$, and $B_\infty$, $\Delta$ is fixed at -174\,G.  The fits are weighted by the inverse uncertainties. The resulting Feshbach resonance parameters are given in Table~\ref{table:parameters}. }
\label{fig:bindingenergies}
\end{figure}

The results of the measurement of binding energy vs.~$B$ are displayed in
Fig.~\ref{fig:bindingenergies}(a). In the universal regime (see Fig.~\ref{fig:bindingenergies}(b)),
where $a$ is much larger than any characteristic length scale of the interaction potential,
$E_\mathrm{b} = \hbar^2/ma^2$, where $m$ is the atomic mass \cite{Duine04, ChinRMP10}.  The solid
lines in Fig.~\ref{fig:bindingenergies} show the results of fitting $E_\mathrm{b}$ in this
universal regime to $a$, where $a$ is given by the usual Feshbach resonance expression
\begin{equation}\label{eqn:fr}
a = a_\mathrm{bg} \left( 1 - \frac{\Delta}{B - B_\infty} \right),
\end{equation}
and where $a_\mathrm{bg}$ is the background scattering length, $\Delta$ is the width of the
resonance, and $B_\infty$ is the location of the resonance. These three quantities are the only fitted parameters.  For $a \gg |a_\mathrm{bg}|$, the
first term in Eq.~(\ref{eqn:fr}) is small, and the fit is insensitive to $a_\mathrm{bg}$ and $\Delta$ separately.  We fix $\Delta = -174$\,G (discussed below) and fit to just two free parameters,
$B_\infty$ and $a_\mathrm{bg}$.  The fitted values are given in Table~\ref{table:parameters}.
In Fig.~\ref{fig:bindingenergies}(c) and (d), the same data is recast in terms of
$\gamma \equiv (mE_\mathrm{b}/\hbar^2)^{1/2}$, where for
large $a$, $\gamma$ vs.~$B$ is approximately linear, as is shown in Fig.~\ref{fig:bindingenergies}(d).
We find no significant difference in $E_\mathrm{b}$ between a BEC or a thermal gas, to within our uncertainties.

Figure \ref{fig:bindingenergies} suggests that the universal regime extends down to $\sim$728\,G,
or $\sim$10\,G below resonance.  Significant discrepancies between the measured $E_\mathrm{b}$  and
universal theory are observed as the field is decreased further.  It is not surprising that the
universal regime spans only a small fraction of $\Delta$, since the $^7$Li resonance is known
to be intermediate between closed-channel and open-channel dominated, as the resonance strength
parameter $s_\mathrm{res} \simeq 0.56$ \cite{footnote} is neither $\ll 1$ nor $\gg 1$
\cite{Kohler06,ChinRMP10}.  For a more precise determination of $a$ it is desirable to extend the
analysis into the non-universal regime, where short range attributes of the potential become
appreciable, and where several of the previously identified Efimov features occur \cite{PollackSci09, Gross10, Gross11}.  A simple two-channel approach to correct for finite range effects, suggested in
Ref.~\cite{ChinRMP10} and applied to $^7$Li in Refs.~\cite{Gross10, Navon11}, is to replace the
universal binding energy expression with
\begin{equation}\label{eqn:binding}
E_\mathrm{b} = \frac{\hbar^2}{m (a - \bar{a} + R^*)^2},
\end{equation}
where $\bar{a} = 31\,a_0$ is the mean scattering length \cite{Gribakin93} (closely related to the van
der Waals radius $a_{vdW} = 32.5\,a_0$), and $R^* = \bar{a}/s_\mathrm{res} = 55\,a_0$ is related to the resonance width \cite{Petrov04}.  The bound state has predominately open channel character only for $a \gg 4R^*$,which is the expected range of validity of Eq.~\ref{eqn:binding} \cite{ChinRMP10}.  The best fit to the data using Eqs.~\ref{eqn:fr} and \ref{eqn:binding} is plotted in Figs.~\ref{fig:bindingenergies}, and the results are presented in Table~\ref{table:parameters}.

\begin{figure}
\includegraphics[angle=-90,width=1\columnwidth]{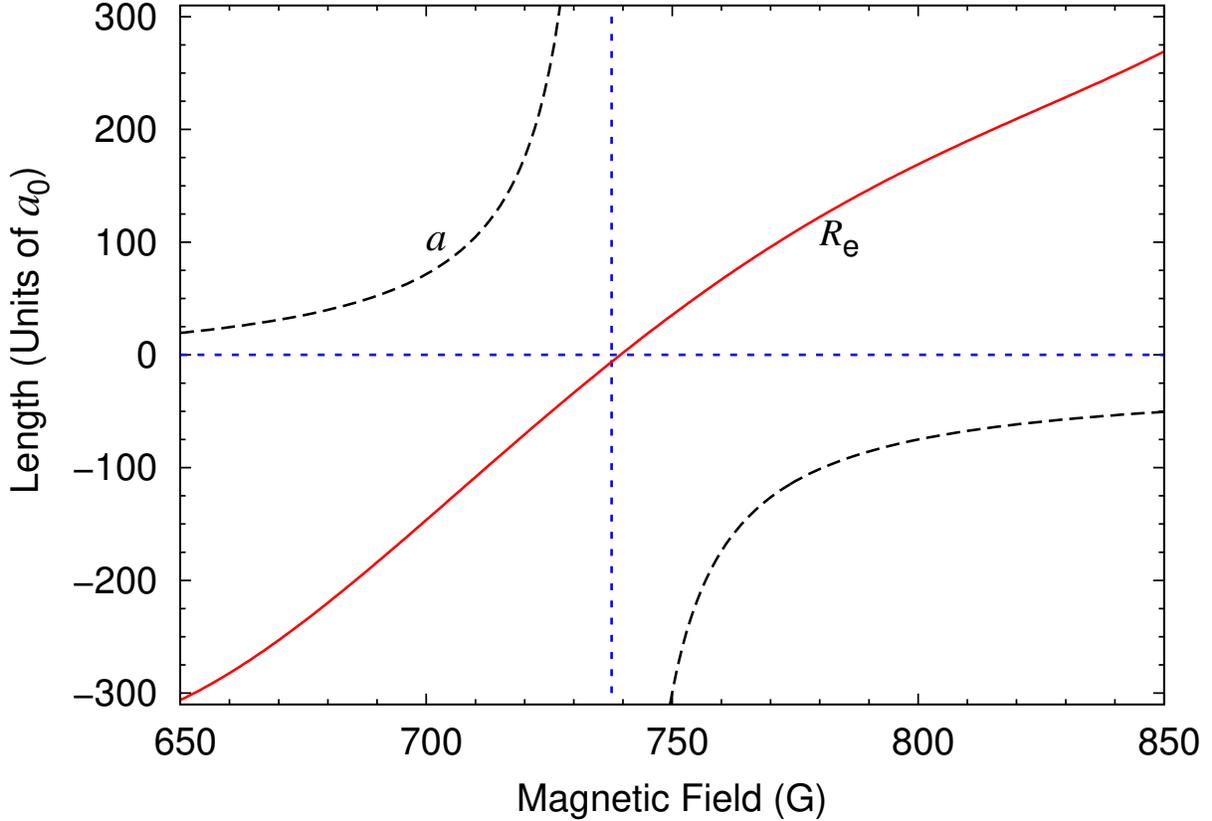}
\caption{(Color online) Coupled channels calculation of $a$ and $R_\mathrm{e}$ for the $F=1, m_F=1$
Feshbach resonance in $^7$Li.  The horizontal and vertical dashed (blue) lines indicate $a=0$ and
$B_\infty$, respectively.  $R_\mathrm{e}$ was fit to a polynomial expansion in the scaled field $\beta
= (B-737.7$\,G)\,/\,G to obtain $R_\mathrm{e}(\beta)/a_0 = -6.2  \,+\, 3.50\,\beta \,-\, 9.2\times 10^{-3}\,\beta^2 \,-\, 6.5\times 10^{-5}\,\beta^3 \,+\, 5.7\times 10^{-7}\,\beta^4$.  This polynomial fit is used to obtain $R_\mathrm{e}$ in Eq.~(\ref{eqn:gamma}). Similar calculations for the $F=1, m_F=0$ \cite{Gross09} and $F=1, m_F=1$ \cite{Ji10} resonances have been previously presented.  In the latter case, the calculation of $R_\mathrm{e}$ agrees well with our results.
\label{fig:FB_RE} }
\end{figure}

Although Eq.~\ref{eqn:binding} gives a somewhat better fit to the data than the universal binding
energy  relation, it clearly fails to represent the entire range of measurements.  This is not unexpected as we are comparing the model to data outside its range of validity.  We find that higher order corrections to this theory offer little improvement to the overall fit quality \cite{Gao04,
ChinRMP10}.  The simple two-channel approach (Eq.~\ref{eqn:binding}) represents the effective range of the potential, $R_\mathrm{e}$, with a single value.   Since the $^7$Li resonance is not open-channel dominated, however, $R_\mathrm{e}$ exhibits considerable field-dependence over the width of the resonance. Properly accounting for this field variation should provide a better correction for finite range effects. In order to obtain $R_\mathrm{e}(B)$ we numerically solved the full coupled-channels equations using realistic model potentials for both the singlet (closed channel) and triplet (open channel) potentials of the electronic ground state of Li \cite{Abraham97, Houbiers98}.  These potentials have been refined by adjusting parameters, such as the potential depth and the shape of the inner wall, to give quantitative agreement with experimentally known quantities, which are primarily the locations of Feshbach resonances \cite{PollackPRL09, Gross11}, zero crossings \cite{PollackPRL09},
and the binding energies of the least bound triplet molecule \cite{Abraham95a}.  The scattering
length and effective range are determined from the energy dependence of the $s$-wave phase shift
$\delta_0$:
\begin{equation}\label{eqn:phaseshift}
k \cot \delta_0 (k) = -\frac{1}{a} + \frac{1}{2} R_\mathrm{e} k^2 + \dots ,
\end{equation}
where $\hbar^2 k^2/m = E_\mathrm{kin}$. Figure \ref{fig:FB_RE} shows both $a$ and $R_\mathrm{e}$
near the Feshbach resonance at 738 G.  There is considerable variation in $R_\mathrm{e}$ over the
width of the resonance, contrary to the assignment $R_\mathrm{e} = -2R^* = -111\,a_0$
\cite{Petrov04, Bruun05}, or $R_\mathrm{e} =2(\bar{a} - R^*) = -49\,a_0$ \cite{ChinRMP10}.  In
comparison, the coupled-channels calculation gives $R_\mathrm{e}(B_\infty) \simeq -6\,a_0$.  Given
this discrepancy, it is not surprising that Eq.~\ref{eqn:binding} does not describe the data well.

More complex solutions to the two-channel model are given in Refs.~\cite{Lange09} and
\cite{Prico11}.  These complex two-channel models improve upon the simple two-channel model by incorporating a field-dependent effective range. The solution to $\gamma$ in Ref.~\cite{Prico11} (Eq.~26) is given in terms of $R^*$ and a short-range parameter $b$, which is related to the van der Waals length and hence, to $\bar{a}$.  This short-range parameter is not universal, but rather is model dependent, and is thus unknown \emph{a priori}.  One way to estimate $b$ is to require that $E_\mathrm{b}$ agrees with Eq.~\ref{eqn:binding} when $1/a = 0$.  In this case, $b = \frac{\sqrt{\pi}}{2} \bar{a}$.  However, we can use the coupled-channels calculation of $R_\mathrm{e}$ to obtain a more informed estimate.  For
$b = 0.85\sqrt{\pi}\, \bar{a}$, the on resonance values of $R_\mathrm{e}$ calculated from the model of Ref.~\cite{Prico11} and from the coupled channels are equal.
Using this value for $b$ and the previously specified values of $R^*$ and $\bar{a}$, the best fit to the data is shown in Figs.~\ref{fig:bindingenergies}.  The expected improvement over the simple model (Eq.~\ref{eqn:binding}) is borne out, as its range of validity (stated as $a \gg a_{vdW}$) is proven to extend to larger detunings from resonance.

The relation between the binding energy of a weakly bound state, or equivalently $\gamma$, and $a$
and $R_\mathrm{e}$ is given by $\gamma = 1/a + \frac{1}{2} R_\mathrm{e} \gamma^2$ \cite{Bethe49}.
Although this quadratic equation has two solutions, only the following has the correct asymptotic behavior for $|R_\mathrm{e}/a| \ll 1$ \cite{Petrov04, Thogersen08, Prico11}:
\begin{equation}\label{eqn:gamma}
\gamma = \frac{1}{R_\mathrm{e}} \left(1 - \sqrt{1 - \frac{2R_\mathrm{e}}{a}}\,\right).
\end{equation}
Figures \ref{fig:bindingenergies} show the results of fitting the data to Eqs.~(\ref{eqn:fr}) and
(\ref{eqn:gamma}), using the $R_\mathrm{e}$ values from Fig.~\ref{fig:FB_RE}.  The agreement
between theory and experiment is very good over a much larger range of the measurements than for
the other models considered, and we therefore use this fit to define the Feshbach parameters, which
are indicated in bold in Table~\ref{table:parameters}.  As previously mentioned, the data is
insufficient to separately extract both $a_\mathrm{bg}$ and $\Delta$ since $a \gg |a_\mathrm{bg}|$.
Given the precise knowledge of the location of the field where $a = 0$, $B_0 = 543.6$, found in our
previous work~\cite{PollackPRL09}, a logical choice would be to fix $\Delta = B_0 - B_\infty =
-194.1$\,G.  The best values of $\Delta$ and $a_\mathrm{bg}$, however, may vary over the large
magnetic field range between the resonance and the zero-crossing.  We find that $\Delta = -174$\,G
gives a slightly better agreement to the coupled channel results for $a > 100\,a_0$, so we adopt
this value. We stress, however, that the fit to the data strongly constrains the product
$a_\mathrm{bg} \, \Delta$, but not each parameter separately.  The differences in $a(B)$ using
Eq.~(\ref{eqn:fr}) with either value of $\Delta$ are less than $2\%$ for $a >100\,a_0$.  A similar
procedure was followed in Refs.~\cite{Gross10, Gross11}. Our Feshbach parameters agree with
Ref.~\cite{Navon11}, where they find $B_\infty = 737.8(2)$\,G, and, although not quite as well,
with Ref.~\cite{Gross11}, which reports $B_\infty = 738.2(4)$\,G.  Finally, we remark that the
model potentials used in the coupled channels calculation are not \emph{a priori} sufficiently
well-known to determine $a(B)$ as accurately as the binding energy data.  The effective range, on
the other hand, varies slowly with $B$ in the region of interest, and we find empirically that its
contribution to the uncertainty in $\gamma$, via Eq.~(\ref{eqn:gamma}), is small.

\begin{table}
\begin{center}
\begin{tabular}{l@{\extracolsep{20pt}}llll}
\hline\hline
Model & $B_\infty$~(G) & $\Delta$~(G) & $a_\mathrm{bg}$~$(a_0)$ & $a_\mathrm{bg}\,\Delta$~(G\,$a_0$) \\
       \hline
Universal
    & 737.82(12) & -174 & -21.0 & 3660(60)\\
Simple two-channel (Eq.~\ref{eqn:binding})~\cite{ChinRMP10}
    & 737.68(12) & -174 & -19.6 & 3410(60)\\
Complex two-channel~\cite{Prico11}
    & 737.73(12) & -174 & -20.4 & 3550(60)\\
Coupled-channels $R_\mathrm{e}$
    & \textbf{737.69(12)} & \textbf{-174} & \textbf{-20.0} & \textbf{3480(60)}\\
\hline\hline
\end{tabular}
\caption{Feshbach resonance parameters obtained by fitting $\gamma$ to Eq.~\ref{eqn:fr}
using the various models.  There are large
uncertainties in $a_\mathrm{bg}$ and $\Delta$ separately, but their product is well-defined by the
data.  The choice of $\Delta$ was guided by the coupled-channels calculation. The quoted
uncertainty in $a_\mathrm{bg}\,\Delta$ reflects shot-to-shot variations in the field and fitting uncertainties.  The uncertainty in $B_\infty$ is systematic uncertainty in field calibration.  Since the range of validity of the non-universal models is guaranteed only for $|R_\mathrm{e}| \ll a$, data
below 725~G is excluded from the fit.  The parameters obtained from the fit to Eq.~(\ref{eqn:gamma}) using $R_\mathrm{e}$($B$) from Fig.~\ref{fig:FB_RE} are our recommended values, and are given in bold. \label{table:parameters}}
\end{center}
\end{table}

\begin{figure}
\includegraphics[width=0.5\linewidth,bb= 220 105 518 507]{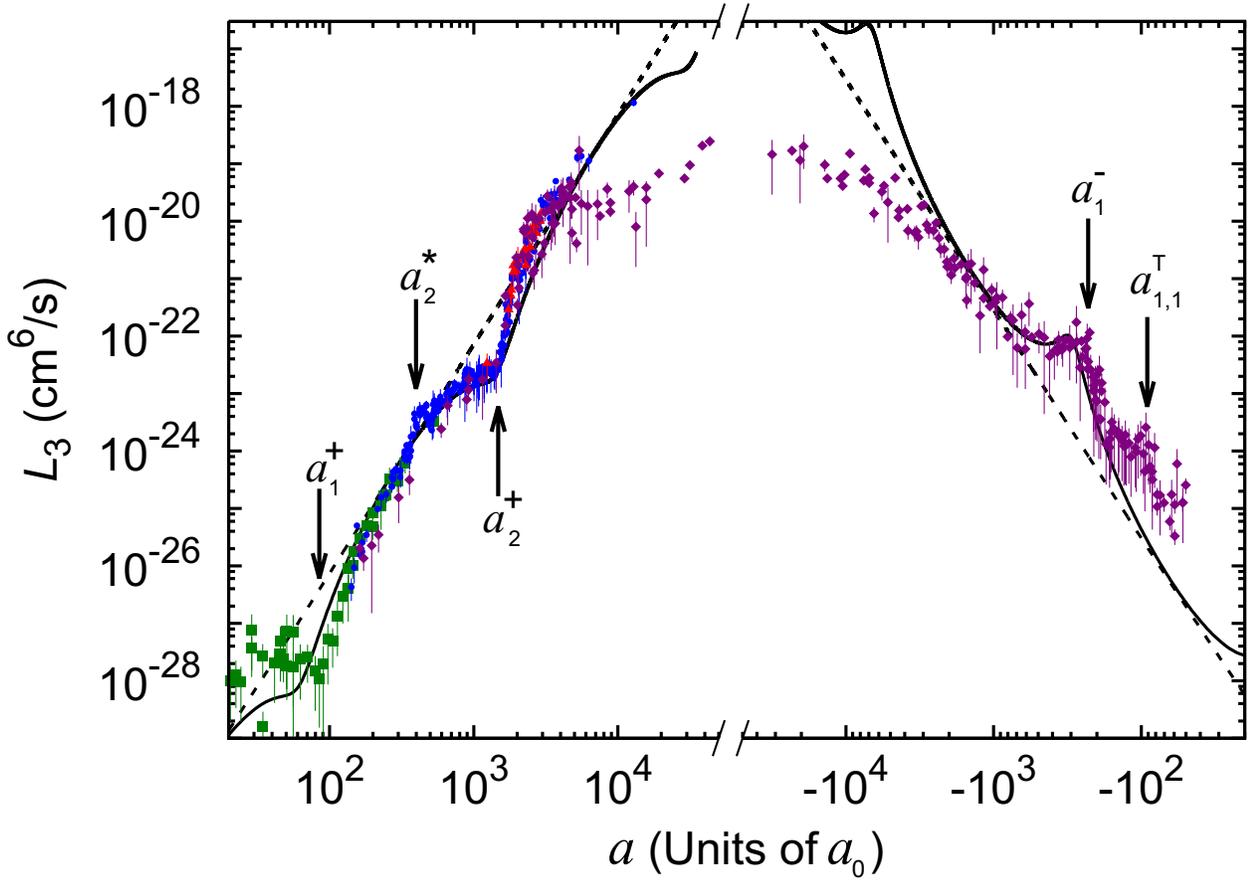}
\caption{(Color online) Three-body loss rate coefficient $L_3$ vs.~$a$. The points are values
extracted from the measured trap loss, with the green, blue, and red points corresponding to
condensates with different trap frequencies, and the purple to a thermal gas (T $\approx$ 1-3
$\mu$K), as reported in Ref.~\cite{PollackSci09}. The solid line shows universal scaling in $a$,
where the positions of the features are determined by the single feature $a^+_2$.  The only
additional fitted parameters are the widths $\eta_+ = 0.075$ and $\eta_- = 0.17$, for the $a>0$ and
$a<0$ sides of the resonance, respectively, and an overall scale factor of 5.5 on the $a>0$ side of
resonance. The need for scaling the universal theory for positive $a$ is unknown.  We ascribe the
deviation from the universal curve for small, negative $a$ to the presence of the four-body feature
$a^T_{1,1}$. The dashed lines are guides to the eye, showing $a^4$ scaling. \label{fig:L3} }
\end{figure}

We now turn our attention to the three- and four-body Efimov features previously reported in
Ref.~\cite{PollackSci09}.  Figure \ref{fig:L3} shows the measured three-body loss rate coefficient
$L_3$ plotted vs.~$a$, where the correspondence between measured values of $B$ is now determined by
the new Feshbach parameters given in Table \ref{table:parameters}.  While $L_3$ generally scales as
$a^4$, as indicated by the dashed lines, it is punctuated by several minima and maxima, which arise
from the presence of Efimov molecular states. The previously reported \cite{PollackSci09} Efimov
maximum $a^-_2$, corresponding to the second Efimov trimer, was an error since the upward shift in
the resonance position by 0.7 G relative to the previous measurement places this feature in the
regime where the loss rates are limited by quantum mechanical unitarity \cite{Weber03, DIncao04,
Rem13}, where $a^-_2$ cannot be resolved in the data.  For the same reason, the effect of the
second Efimov tetramer associated with the second trimer ($a^T_{2,2}$) is also not visible in
measurements of the four-body loss rate coefficient $L_4$ (not shown).  The fitted locations of the
remaining features are given in the second column of Table \ref{table:Efimov}, where the caption
provides a key to the notation.  Not all of the features given in Table \ref{table:Efimov} are
indicated in Figure \ref{fig:L3}, but expanded views of both $L_3$ and $L_4$ are found in
Ref.~\cite{PollackSci09}.

%

\vspace{\baselineskip}

\begin{table}{}
\begin{tabular}{@{\extracolsep{18pt}}p{30pt}cccc}
    & Feature & Experiment ($a_0$) & Universal Scaling in $a$ ($a_0$) & \\
\hline
$a > 0$
    & $a^+_1$   & $89(4)$       & $62.57^*$  &\\[3pt]
    & $a^+_2$   & $1420(100)$    & $(1420)$   &\\[3pt]
    & $\left[a^*_2\right]$   & $421(20)$     & $317.5^{*\dag\ddag}$ &\\
    & $\left[ a^*_{2,1}\right]$ & $912(50)$ & $697.3^\S$ &\\[3pt]
    & $\left[ a^*_{2,2}\right]$ & $1914(200)$ & $2154^\S$  &\\
 \\
$a < 0$ & $a^-_1$   & $-252(10)$ & $-298.1^\dag$   &\\[3pt]
    & $a^-_2$       & ---       & $-6765^\dag$   &\\[3pt]
    & $a^T_{1,1}$   & $-94(4)$     & $-126.8^\sharp$    &\\
    & $a^T_{1,2}$   & $-236(10)$    & $-272.0^\sharp$   &\\[3pt]
    & $a^T_{2,1}$   & $-4060(800)$   & $-2878^\sharp$   &\\[3pt]
    & $a^T_{2,2}$   & ---       & $-6173^\sharp$   &\\[3pt]
\hline
\multicolumn{5}{l}{
{\footnotesize References:
    $^*$ \cite{Braaten06};
    $^\dag$ \cite{Helfrich10};
    $^\ddag$ \cite{Gogolin08};
    $^\S$ \cite{Deltuva11};
    $^\sharp$ \cite{Deltuva12}.
    }
    }
\end{tabular}
\caption{Locations of Efimov features, given in units of $a_0$, of the three- ($L_3$) and four-body
($L_4$) loss coefficients. The experimental values of $a$ are extracted from the measured fields
using Eq.~(\ref{eqn:fr}) with the parameters in bold from Table \ref{table:parameters}. The horizontal lines indicate features that are near the resonance and were not observed. The estimated uncertainties include fitting uncertainties, as well as the
uncertainties in the Feshbach parameters. For $a > 0$, $a^+_i$ denotes the recombination minimum of
the i$^\mathrm{th}$ Efimov trimer.  The origins of three features, labeled as $\left[a^*_2\right]$,
$\left[ a^*_{2,1}\right]$, and $\left[ a^*_{2,2}\right]$, have not been identified, but roughly
correspond to expected locations of atom-dimer and dimer-dimer resonances. For $a<0$, $a^-_i$
denotes the Efimov resonance where the i$^\mathrm{th}$ trimer merges with the free atom continuum.
The remaining features, $a^T_{i,j}$, arise where the j$^\mathrm{th}$ tetramer associated with the
i$^\mathrm{th}$ trimer merges with the free atom continuum.  The final column gives the
predicted locations of the features using universal scaling in $a$. The
universal scaling relations were obtained from the indicated references. The scaling is anchored by
the measured location of $a^+_2$, which, as an input, is denoted by parentheses.}

\label{table:Efimov}
\end{table}

The origins of three of the features in Table \ref{table:Efimov}, indicated by square brackets, are
uncertain. The feature $a^*_2$ is nominally located at the atom-dimer resonance where the energy of
the second Efimov trimer merges with the atom-dimer continuum.  Relatively sharp peaks in $L_3$,
located near the expected atom-dimer resonance, were previously reported  for $^{39}$K
\cite{Zaccanti09}, and $^7$Li \cite{PollackSci09, Machtey12}. Since a large dimer fraction is
unexpected, a model was developed to explain the presence of enhanced loss even without a large
population of dimers \cite{Zaccanti09}.  In this model, each dimer produced in a three-body
recombination collision shares its binding energy with multiple atoms as it leaves the trap volume
due to the enhanced atom-dimer cross section \cite{Zaccanti09}.  Recent Monte Carlo calculations,
however, conclude that the resulting peak from this avalanche mechanism is too broad and shifted to
higher fields to explain the observations \cite{Langmack12}. The remaining two features, $a^*_{2,1}$ and $a^*_{2,2}$, are nominally located at dimer-dimer resonances, where the energy of a tetramer merges with the dimer-dimer threshold \cite{DIncao09}. Their assignment also remains tentative, since
their observation requires a significant and unsubstantiated dimer population.

The third column in Table \ref{table:Efimov} gives the predictions of universal scaling.  Many of
the scaling relations presented in the pioneering papers for the three-body \cite{Braaten06} and
four-body sectors \cite{vonStecher09, DIncao09}, have been replaced by the more precise theoretical
determinations cited in Table \ref{table:Efimov}. Four significant digits are given to reflect the
stated precision of these scaling relations. If the relative positions of all features are
universally connected, the position of only one is needed to completely fix the remaining.  We
choose the recombination minimum of the second trimer, $a^+_2$, for purpose of comparison, as it is
a well-defined feature that occurs at sufficiently large $a$ ($\sim$$1400\,a_0$) to be insensitive
to short-range effects, while also being small enough in magnitude to not be hypersensitive to $B$.
While the measured locations are consistent with universal theory at the 20-30\% level, some of the
features, in particular $a^+_1$ and the lowest tetramer $a^T_{1,1}$, occur deep in the
non-universal regime where $|R_\mathrm{e}/a| > 1$.  We attempted to correct the universal theory
for the effect of finite range using the same strategy applied to the dimer binding energy, that is
by applying universal scaling in $\gamma^{-1}$ (Eq.~\ref{eqn:gamma}) rather than in $a$.  To lowest
order, the correction to $1/a$ is $\frac{1}{2} R_\mathrm{e}/a^2$.  We find that such a replacement
improves the agreement with experiment for features on the $a<0$ side of the resonance, but for
$a>0$ the agreement is actually made worse.
We note that an effective field theory for short range interactions has
been developed in which corrections to universal scaling of three-body quantities are computed to
$\mathcal{O}$($R_\mathrm{e}$) and that they have been applied to the $F=1, m_F=0$ Feshbach
resonance in $^7$Li \cite{Ji10}.  Effective range corrections have also been used to analyze Efimov
features in Cs \cite{Hadizadeh13}.  It would be interesting to apply the same analysis to the $F=1,
m_F=1$ resonance in $^7$Li to compare with the data presented here.

A measure of universality across the Feshbach resonance may be obtained by evaluating the ratio
$a^+_2$\,/\,$a^-_1$.  Universal scaling in $a$ implies $a^+_2$\,/\,$a^-_1 = -4.76$
\cite{Helfrich10}, whereas experimentally, we find $a^+_2$\,/\,$a^-_1 = -5.63$. We disagree with a
previous measurement for the $|F=1, m_F=1\rangle$ state in $^7$Li, which found $a^+_2$\,/\,$a^-_1 =
-4.61$ \cite{Gross11, footnoteGross}.  The Efimov features observed in
Refs.~\cite{Gross10, Gross11} are not as sharp as those reported here, and this may affect the
precision for which the location of a feature is extracted.  The width is quantified by the fit
parameter $\eta$, which is related to the lifetime of the Efimov molecule \cite{Braaten06}.  For
Ref.~\cite{Gross11}, $\eta_+ = 0.17$ and $\eta_- = 0.25$, corresponding to the $a>0$ and $a<0$
sides of resonance, respectively, while we find $\eta_+ = 0.075$ and $\eta_- = 0.17$.  These large differences, at least in case of $\eta_+$, may indicate that $\eta$, and hence the dimer lifetime, has an interesting and unexpected temperature dependence, since the $a>0$ data in
Refs.~\cite{Gross10, Gross11} is obtained with a thermal gas, while in our experiment the gas is
cooled to nearly a pure Bose condensate.

It was pointed out recently that the location of the first Efimov trimer resonance $a^-_1$, when
scaled by the van der Waals radius $a_{vdW}$, is remarkably similar for multiple unconnected
resonances in the same atom \cite{Berninger11}, as well as for different atomic species
\cite{Chin11}.  These observations suggest that there is no need for an additional ``three-body
parameter" to pin down the absolute positions of the Efimov features, but rather, that this scale
is set by short-range two-body physics \cite{Chin11, Wang12, Schmidt12, Sorenson12, Knoop12,
Naidon12}.  For the measurements reported here, -$a^-_1$\,/\,$a_{vdW} = 7.8$, which is
close to the range of 8-10 reported in most other cases
\cite{Berninger11,Chin11,Knoop12,Roy13}.

Quantum mechanical unitarity implies that $L_3$ is limited for non-zero temperatures, as is evidenced by the purple points in Fig.~\ref{fig:L3} near the resonance, for which the highest average $L_3$ is $\sim$$8\times10^{-20}$\,cm$^6$/s.  This value is about 3 times greater than the largest $L_3$ \cite{Rem13} predicted for a $1\, \mu$K thermal gas, which is the lowest temperature of our thermal data \cite{PollackSci09}.  This discrepancy may indicate a systematic error in measuring $L_3$ under conditions where the decay rate is comparable to the rate of thermalization.

The determination of the Feshbach parameters for $^7$Li in the $|F=1, m_F=1\rangle$ state by direct
measurement of the dimer binding energy is a significant improvement over our previous measurement
using condensate size. We have measured the dimer binding energy deep into the non-universal regime
and find that data are well-represented by corrections based on the field-dependent effective
range. Using these more precise parameters we find that the overall consistency between the
experimentally determined locations of three- and four-body Efimov features and those obtained from
universal scaling is in the range of 20-30\%.  Since we use the location of only one feature as
input, the agreement supports the contention of universal scaling across the Feshbach
resonance, but also points to the need for a better understanding of effective range corrections to
Efimov spectrum. The origin of features nominally located at atom-dimer and dimer-dimer resonances
remains an open question.  A direct measurement of the equilibrium dimer fraction could help to
resolve this issue~\cite{Langmack13}.

We thank Eric Braaten, Georg Bruun, Cheng Chin, and Paul Julienne for helpful discussions.  Funding was provided by the NSF, ONR, and the Welch
Foundation (Grant C-1133), and the Texas Norman Hackerman
Advanced Resources Program.

\begin{thebibliography}{10}
\bibitem{footnote}
The value of $R^*$ differs from that given in Ref.~\cite{ChinRMP10} due the change in Feshbach
resonance parameters compared with those given in Ref.~\cite{PollackSci09}.
\bibitem{footnoteGross}
The notation, $a_+$ and $a_-$, in Refs.~\cite{Gross10,Gross11} differs from ours: $a^+_2 =
4.764\,a_+$ and $a^-_1 = a_-$.   Furthermore, the universal ratio given as $a_+$/$a_- = 0.96$ in
those references should be updated to 1.0, according to Ref.~\cite{Helfrich10}.
\end{thebibliography}
\hyphenation{Post-Script Sprin-ger}

\end{document}